\documentclass[conference]{IEEEtran}
\usepackage{hyperref}
\usepackage{graphicx}

\begin{document}

\title{Experimental Study of Application Specific Source Coding for Wireless Sensor Networks}

\author{
\authorblockN{ Muthiah Annamalai}
\authorblockA{ 
 muthiah.annamalai@uta.edu}
\and
\authorblockN{ Darshan Shrestha}
\authorblockA{
darshan.shrestha@uta.edu\\
\\
Department of Electrical Engineering,\\
University of Texas at Arlington, USA.\\
}
\and
\authorblockN{ Saibun Tjuatja,}
\authorblockA{ tjuatja@uta.edu}
}
\maketitle

\begin{abstract}
The energy bottleneck in Wireless Sensor Network(WSN) can be
 reduced by limiting communication overhead. Application specific
 source coding schemes for the sensor networks provide fewer bits to
 represent the same amount of information exploiting the redundancy present
  in the source model, network architecture and the physical process.
  This paper reports the performance of representative codes from
  various families of source coding schemes (lossless, lossy, constant bit-rate,
variable bit-rate, distributed and joint encoding/decoding ) in terms of 
 energy consumed, bit-rate achieved, quantization-error/reconstruction-error, 
latency and complexity of encoder-decoder(codec). A reusable frame work for
 testing source codes is provided. Finally we propose a set of possible 
applications and suitable source codes  in  terms of these parameters.
\footnote{ This work was done at Wave Scattering Research Center during April-May 2006. }
\footnote{The source code for the TinyOS implementation of various codecs
can be obtained freely upon request to the primary author.}
\end{abstract}

\IEEEpeerreviewmaketitle

\section{Introduction}
WSNs exhibit new approaches to providing reliable, time-critical and
constant environment sensing, event detecting and reporting, target
localization, and tracking. Since the power supply cannot be 
replenished in most cases, energy 
efficient techniques and protocols are needed to extend the lifetime
of the system. 

Generally cost of transmission is higher than
the cost of processing  \cite{chandrakasan_amps}. Communicating the
data after source coding \cite{text_book},  can save energy by reducing the
number of bits transmitted over radio. Similarly, channel coding  can
provide a better Bit Error Rate(BER) than achieved at the uncoded
power level.  This coding gain represents energy savings. Local
processing is useful as long as it leads to energy savings compared to
uncoded transmission  \cite{chandrakasan_amps}.

In this paper, source coding techniques that exploit statistical 
redundancy in sensor
data and correlations in space and time were used. In dense sensor
networks, observed functions of physical environments change slowly
over small distances. This sensor data are correlated in
space. Moreover these physical processes do not change abruptly in
time, so the readings are correlated in time as well. This redundancy
due to temporal and spatial correlation can be exploited. Such
correlation structures allow sophisticated compression and distributed 
source coding schemes from Information Theory(eg. DISCUS) \cite{DISCUS},
Signal Processing(eg. Haar Wavelet) \cite{code_haar} to be used.

Source coding techniques for correlated data seek to translate
the theoretical coding-rate of Slepian-Wolf limits
\cite{SlepianWolf} to practical codes. Recently the DISCUS approach has generated new 
techniques to construct source codes to reach this limit. The other idea
in correlated data compression comes from signal
processing, which is the Haar wavelet \cite{code_haar}. 

Sensor networks are application specific by design. So various
protocols and architecture  used have to be customized
for each network. For example, a sensor network deployed in the hilly
 mountains for environmental monitoring do not have stringent
 requirement on event detection compared to sensor network deployed
 for intruder detection. Dropped packets and inaccuracy may not
 be detrimental in the case of environmental monitoring but same
 amount of low-accuracy and high-delay might defeat the purpose
 of a intruder detection network. Therefore, it is up to an
 application to somehow define the degree of accuracy of the results
required. Source coding models depend largely on the tolerance
of each network towards approximated results (lossless,lossy), latency of
coder-decoder and energy consumed. This requires source codes which are
application specific.
The contributions of the paper are heuristic mapping
of source codes with possible applications, and a reusable frame work for 
testing source codes in WSNs.
The following sections discuss coding schemes, experimental setup, implementation
details and results.
\section{Source Codes}
%
%
The characteristics of various source codes 
 are reviewed. The source codes can be broadly classified according to Table 1. 

\begin{table}[h]
\caption{Source Code Classification}
\begin{tabular}{|l|l|l|}\hline
 Parameter &  \multicolumn{2}{|c|}{ Code}  \\ \hline
 Loss & lossless & lossy \\ \hline
      & T-code,Fibonacci  &  DISCUS,[A\&$\mu$]-law,\\
      & & Modulo-code \\ \hline
 Joint & coding \& decoding &  decoding \\ \hline
 & Haar,Modulo-code & DISCUS \\ \hline
 Bit rate & Variable & Constant\\ \hline
  & Fibonacci, T-Code & $\mu$-law, A-law, DISCUS \\ \hline
\end{tabular}
\end{table}

\subsection {A-law compander}
A compander is a pair of compressor and expander functions which 
perform in tandem, generally in lossy coding. The A-law compander 
is a non-uniform quantization function which resolves linearly
 upto certain significant values and quantizes the remaining values
 non-linearly. A-law compresser and expander together form a lossy
 compression system that is used to compress the samples of the
 signals \cite{text_comm}. The standard European and widely accepted
 compander is A-law . In this implementation integer-quantized A-law
 scaled to range 0-255 was implemented. A=87.6 was used.

\subsection {$\mu$-law compander}
The $\mu$-law compander was also used in a similar scaled and integer-quantized 
manner like the A-law. $\mu$-law is the accepted standard in the  North America \&
Japan \cite{text_comm}. $\mu$=255 was used.

\subsection{Differential Pulse Code Modulation (DPCM)}
DPCM 
sends first sample in a frame uncoded, while the rest of the 
samples are coded as difference from the previous sample and transmitted. 
At the decoding side, the first sample is taken as such and successive samples
are added to the previously decoded value to generate the decoded samples.

\subsection {Fibonacci code}
It is a form of universal code employed in data compression
with useful property that all codes end with $11$, and no code ends 
with $10$. This implies that even if there are errors in the stream,
it is possible to resynchronize with the data. 

Fibonacci code is essentially a number system based on the weights of the Fibonacci series.
Any number is decomposed as the sum of the weights of Fibonacci numbers. For example, the
value $10$ is encoded into Fibonacci number as $(010011)_2$, which is interpreted as
$10=2.1+8.1$ due to position of $1$ being at the weights $2,8$.

Construction of Fibonacci code can be 
found in \cite{code_fibo}. The variable length code so obtained, can be used to map to symbols based on their probabilities to reduce the average code word length $L_{avg}=\sum_{i=1}^{N}l_ip_i$, where $p_i$ is probability of $i^{th}$ symbol and $l_i$ its code-length in bits, for a $N$ symbol
source.

\subsection {T-code }
T-code \cite{code_t} is a variant of Huffman code built iteratively from the case of 
a simple extension of Huffman code. Once the code size reaches the desired amount of
symbols, the most frequent symbols are assigned code words with the least 
length such that the average code word length ($L_{avg}$) is minimized. 
For example, T-codes are built from the set $S(0)=\{1,00,01\}$, for level-1.
For the next level the any one symbol is prefixed to the rest of the symbols,
to obtain $S(0,1)=\{00,01,11,100,101\}$. Extending to level-3 $S(0,1,00)=\{01,11,100,
101,0000,0001,0011,00100,00101\}$ and so on.

T-codes also have a useful synchronization property that if certain bits are lost
 in the stream, the T-code can self-synchronize unlike other codes which will
 invalidate the whole stream. T-code has good error recovery characteristics
 \cite{code_t2} in addition to being a source-coding algorithm.

\subsection {Correlated Source Coding}
A family of codes specifically designed to exploit the spatio-temporal correlations in the 
sensor data to achieve a lower bit-rate closer to the rate predicted by
Slepian-wolf theorem \cite{SlepianWolf} were chosen and evaluated. 
According to the Slepian-Wolf if X,Y are two correlated sources  a source coding bit-rate of
$R(X,Y)$, such that $H(X)+H(Y) \ge R(X,Y) \ge H(X) + H(Y|X) = H(Y)+ H(X|Y) = H(X,Y)$  \\
is achievable called as the Slepian-Wolf limit for the correlated source coding
schemes.

\subsubsection {Modulo code}
Modulo code uses modulo coding for either the odd or even
nodes. For example, one node [say odd] encoder transmits the 
 modulo-N value of data, and for other node[even] the encoder transmits original data.
At the receiver, decoder computes the \textit{closest} multiple of N, which 
on adding with the residue from odd node, approaches as close as possible to the 
data from the even node. Generally $n=2^k$ or $n=8$ for practical purposes.

Error-free decoding is achievable if and only if data from 2 sources lie in the 
same $8.k$ bin. Example $40,44$ can be encoded as $40,4$ and decoded correctly.
But $40,35$ are encoded as $40,5$ are decoded wrongly as $40,37$. This correlation model
is restricted to cases where the data from 2 sources lie in the same bin.

\subsubsection {Haar Wavelet codes}
Integer Haar wavelet \cite{code_haar} transform models correlated data between the 
two sources by computing the low-pass and high-pass coefficients of the data. 
This scheme in its simplest form was used in the experiment as a 1-level Integer Haar transform
by computing the sum \& difference of the data from the two nodes
which correspond to the LP \& HP Integer Haar coefficients. This is a
type of joint-encoding technique which can also be used in
hierarchical sensor networks for scalable and lossless transforms
with low bit rates, as shown in \cite{code_haar}.

\subsubsection {DISCUS}
Distributed source coding using syndromes \cite{DISCUS_2} is a generic
technique to design source codes that reach the Slepian-Wolf
limit. The basic premise is to model the correlation between sources
as a type of channel noise and then choose a particular type of
partitioned channel code for source coding. If the channel code can
tolerate and perform well against this type of noise, then it is argued in
\cite{DISCUS_Hamming}, that the corresponding source code will perform
well against the correlated data, treated as noise. The correlation between
the two data sources is measured in terms of the Hamming distance of their
data, which need to be $H_{dist} \le t$, where $t$ is the error-correcting
capacity of the code. 

In this implementation, the standard  (7,4) Hamming code was split
into two sub-codes and distributed encoding  was done. The correlation
model can tolerate the samples to be off by at-most $t=1$ bits; for larger
Hamming distance between the two source codes are the sources 
errors in this scheme. By this 7 bits of information were
represented as just 5 bits owing to the correlation model assumed.
The two 5 bit samples were decoded jointly at the decoder and
their respective 7 bit versions were produced with a tolerable
channel error upto 1 bit. This model is the simplest possible of
the DISCUS codes achieved in practice. The DISCUS implementation
follows description of \cite{DISCUS_Hamming}.

\section{Experimental Setup}

\subsection{Network}

\label{fig_nwarch}
\begin{figure}
\centering
\includegraphics[height=1.8in,width=2.0in]{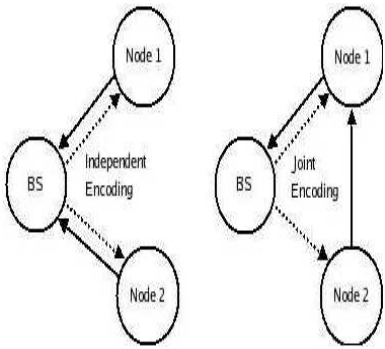}
\caption{Network architecture}
\end{figure}

3 MICA2 motes, arranged in either of the configurations shown in [Fig 1],were used. 

\subsection{Hardware}
In this paper, the MICA2 motes in the RF frequency band 433MHz
(MPR410) were used for experiment purposes. The Motes use the Chipcon
CC1000, FSK modulated radio and Atmega 128L micro-controller. 


\subsection{Data collection}
The data for the experiment were collected from the
photo sensor by walking randomly around the lab, and
varying light intensity by turning on/off the lights, or varying the
window blind positions and gradually changing  the light and
shade  to adjust light intensities on the photo sensor.

For pseudo-sensor codes, the data itself was simulated from the MICA mote,
so a sensor board was not needed.

\subsection{Operation}
The base station was turned on after the  two sensor nodes.
The base station was connected to the PC via UART. After decoding, base station forwarded the 
data received from the sensor nodes to the PC. The data was read from PC serial port
and saved into a log file for performance analysis.

\section{Implementation}
The experiment was performed at a sampling rate of 2Hz, and tested up
to 125Hz for two nodes. No scaling problems or dropped packets were
observed in this scaling of sampling rates. Encoder and decoder
latency was observed to be around the same values at various sampling
rates, as reported.

\subsection{Base station}
The base station was designed conceptually as shown in [Fig 2].
It decoded the received packets, logged it to the PC and checked if
 the sensor nodes needed to be resynchronized and broadcasted the message.

\label{base_arch}
\begin{figure}
\centering
\includegraphics[height=2.6in,width=2.2in]{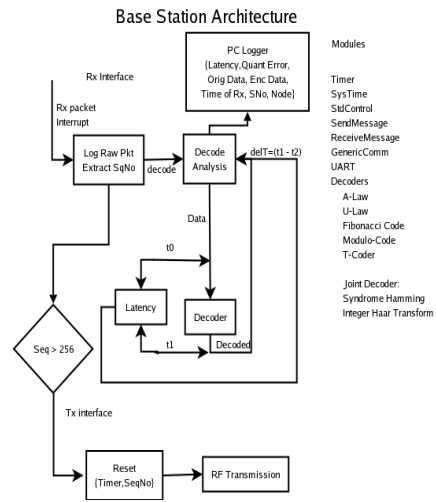}
\caption{Base station architecture}
\end{figure}

\subsection{Sensor Node}
The sensor node [Fig 3] periodically sent out the encoded ADC data to the base station. 
It also resynchronized itself on receiving any packet from the base station.

\label{sense_arch}
\begin{figure}
\centering
\includegraphics[height=2.6in,width=2.2in]{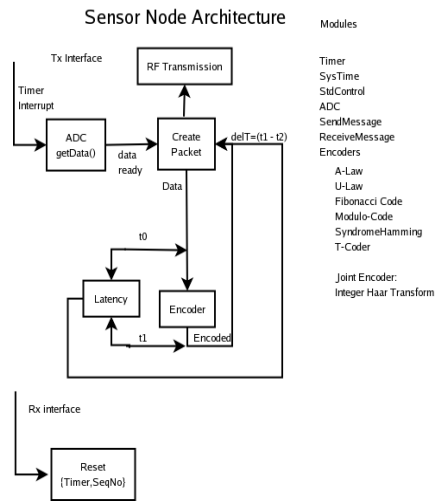}
\caption{Sensor architecture}
\end{figure}

\subsection{Communication}

Immediately after the base station was turned on, a broadcast message was sent 
to reset the sensor nodes and the base station waited for the
data packets. Sensor nodes were programmed to access the channel
in a mutually exclusive manner and sent their encoded samples to the base station.
Once the sequence number reached 255, the base station resent the broadcast
message and the whole process was repeated.

\subsubsection{Packet formats}

\begin{table}[h]
\centering
\caption{Packet Structures}

\label{tbl:txpkt}
\label{tbl:pcdatapkt}
\label{tbl:bcastpkt}

\begin{tabular}{|l|l|l|l|l|l|}\hline
\multicolumn{2}{|c|}{Sensor Data} & \multicolumn{2}{|c|}{PC Data} &
\multicolumn{2}{|c|}{ Broadcast Data} \\ \hline
Field & Byte & Field & Byte & Field & Byte\\ \hline
NodeID & 1 & SensorData Pkt & 9 & Command & 2\\ \hline
Sequence\# & 1 & decodeData & 2 & Timer & 2\\ \hline
codeData & 2 & decodeLatency & 2 & Sensor & 1\\ \hline
originalData & 2 & receiveTime & 4 &  & \\ \hline
Length & 1 & overflow & 2 & &  \\ \hline
Latency & 2 & & & & \\ \hline
\end{tabular}
\end{table}

\label{txpkt}
Transmitter packet format from sensor to the base station is in
Table 1. Most of the fields are self descriptive. Length
field represents the number of bits of the coded data. Latency is
the time taken for the encoding operation, in microseconds. Original
data is the MSB 8 bits of the 10 bit ADC samples.

Base station broadcast reset message was done via a packet
structure shown in  Table 1. Command field was used to
request an action to be performed at the sensor node. Timer interval was the
global time to which all sensors must synchronize with. Sensor field
represents the type of sensor data requested; photo sensor was used
in our implementation. Command and Sensor fields can be extended to 
support various commands.

The base station decoded the packets from the Transmitter sensor,
appended extra fields and forwarded the packet to the PC via the UART 
using a packet structure shown in Table 1. The extra
fields concatenated with the Transmitter packet are again self
explanatory. All the time units are in microseconds. Overflow exists
to count the number of dropped packets at high sampling
rates.

\subsubsection{Channel Access}

\label{fig:mutextimer}
\begin{figure}
\centering
\includegraphics[height=1.6in,width=2.6in]{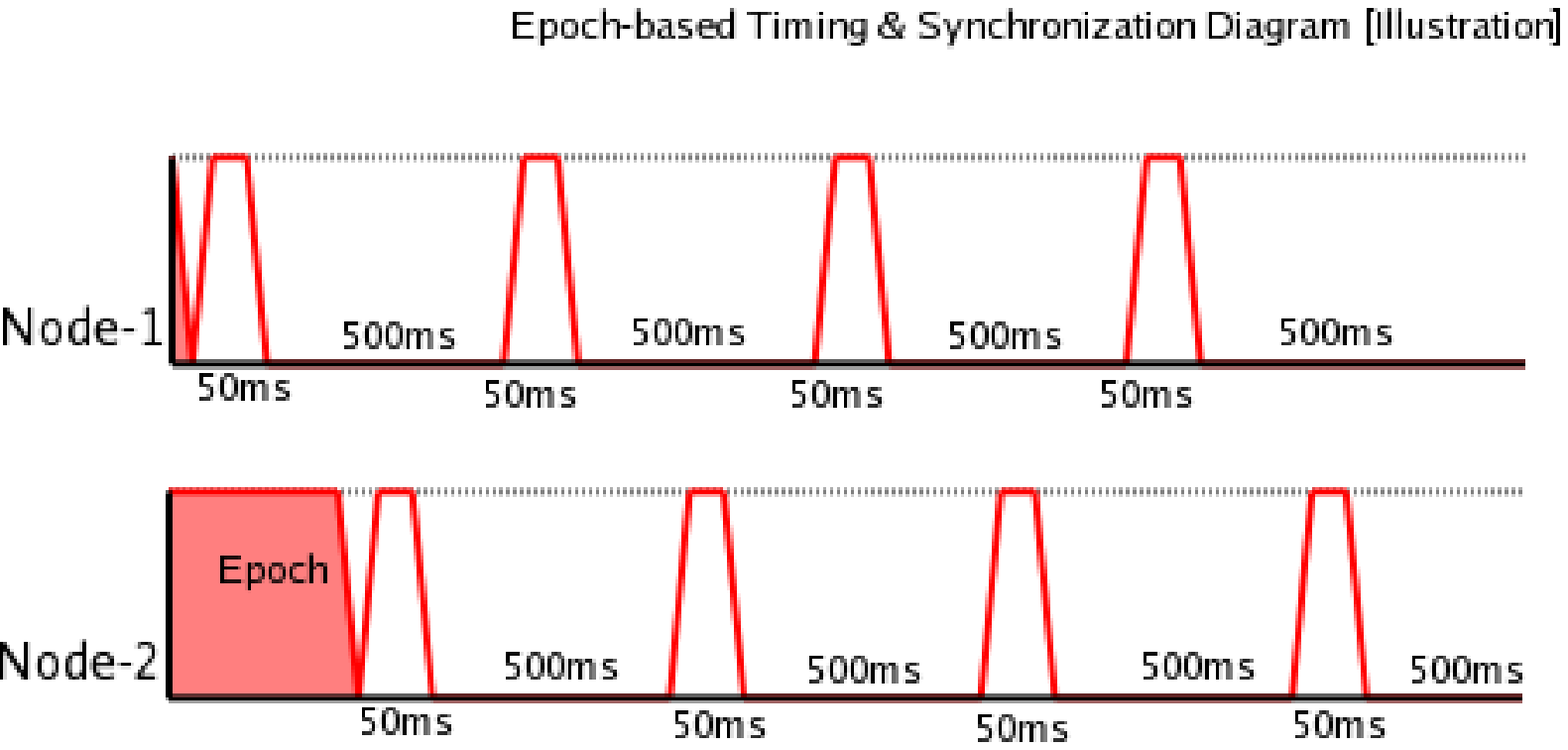}
\caption{Time multiplexing between two nodes.}
\end{figure}

 Scheduling of the sensor nodes was done by allocating fixed time
 slots in a mutually exclusive manner, as in 
[Fig 4]. This is a simplified version of TDMA based
 MAC protocol.

\subsection{Code}
The application was written in NesC on the TinyOS system. Also
various tools were used in the experimental setup:

\subsubsection{Statistics}

%
%

\begin{table}[h]
\centering
\caption{Code Statistics}
\begin{tabular}{|c|c|c|} \hline
 Lines of Code & Language & Description \\ \hline
 $\ge$3000 & NesC & implementation \\ \hline
 $\ge$200 &C&testing \\ \hline
 $\ge$1000& Octave& analysis \& plotting \\ \hline
 $\ge$100& Python& conversion,build-process \\ \hline
 $\ge$100& Java& serial I/O formatting \\ \hline
\end{tabular}
\end{table}
The source code for the entire application is available at \cite{code_download}.

\begin{table}[h]
\centering
\caption{Implementation Code Size}
\begin{tabular}{|c|c|c|c|c|} \hline
 Code Type & \multicolumn{2}{|c|}{ ROM [bytes] } & \multicolumn{2}{|c|}{ RAM [bytes]}\\ \hline
  & B & S & B & S \\ \hline

 $\mu$-law & 0 & 0 & 314 & 280 \\ \hline
 A-law     & 0 & 0 & 314 & 280 \\ \hline
 Fibonacci & 0 & 0 & 586 & 544 \\ \hline
 Fibonacci & 0 & 4 & 584 & 1328 \\ 
 (pseudo)  &   &   &    &  \\ \hline
 Modulo-8  & 1 & 0 & 64 & 14 \\ \hline
 T-code   & 0 & 0 & 546 & 540 \\ \hline
 T-code   & 0 & 4 & 546 & 1190 \\ 
 (pseudo) &   &   & & \\ \hline
 Haar  & 0 & 3 & 82 & 512 \\ \hline 
 DISCUS & 2 & 0 & 1238 & 1292 \\ \hline
 DPCM & 3 & 2 & 36 & 40 \\ \hline 
\end{tabular}
\end{table}

Implementation statistics for each of the source coders are given
as increments to the reference implementation overhead (communication, startup etc)
which occupies (excluding the given source coder) for base station 693b ROM, and
10270b RAM; and for the sensor node 452b ROM, and 10406b RAM. 

\subsubsection{Testing}
All the source codecs [coder-decoder pairs] were written in a 
high-level language  Octave \cite{octave} and tested. 
Next they are manually converted to C code which were 
subsequently converted into NesC. The unit-testing of each codec was 
done before the integration into the system and verified for its accuracy.

\subsection{Codec Implementation}
\subsubsection{Encoder}
The encoders for the codes A-law, $\mu$-law, Fibonacci code, 
T-code [variant of Huffman code] were implemented 
as a Lookup Table(LUT). This basically creates a $O(1)$
time complexity encoder for all these cases with $O(n)$ storage space.

For the other encoders including Modulo code, Haar Wavelet code, 
DISCUS (variant of Hamming code),
a similar constant-time encoder (Modulo,Haar) and linear-time encoder(DISCUS)
were achieved using computation and not LUTs.

\subsubsection{Decoder}
The decoders for A-law, $\mu$-law, Fibonacci code can be decoded by using a
binary-search algorithm on the same LUT as the encoder with a 
known complexity of $O(log(n))$. A $O(n)$ decoder has been implemented for the T-code using a LUT
linear search. As the T-code LUT is built from frequency dependent
data which are  not always sorted, a binary-search algorithm cannot be used.
 For case of Modulo-code, Haar
Wavelet and DISCUS-Hamming code, the decoding complexity is
proportional to $O(n)$, for the Modulo-n case, $O(1)$ as it is just
sum \& difference, and $O(kn^2)$ as it involves a binary-matrix multiplication respectively.

\subsection{Pseudo Data Sources}
 The entropy codes and universal codes  were modeled 
with probability distributions different from the existing data source due to which 
 the average bit rates were not optimal and in some cases (Fibonacci) caused expansion instead
of compression.
To overcome this problem, a deterministic pseudo-data source was designed in
the TinyOS platform in place of the photo sensor. The resulting
probability distributions from these pseudo data sensors were used to 
re-design the codes LUT for T-code, Fibonacci codes. Codecs performing with pseudo-data sources 
were found to yield much better results closer to the optimum average bits/sample $L_{avg}$.

\section{Results}

\begin{figure}
\includegraphics[height=2.5in,width=3in]{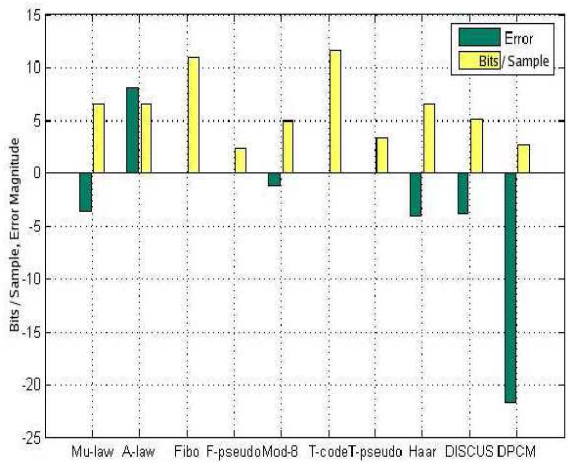}
\caption{Error Rate \& Average bit rate measures for various codes}
\end{figure}

\begin{figure}
\includegraphics[height=3.0in,width=3in]{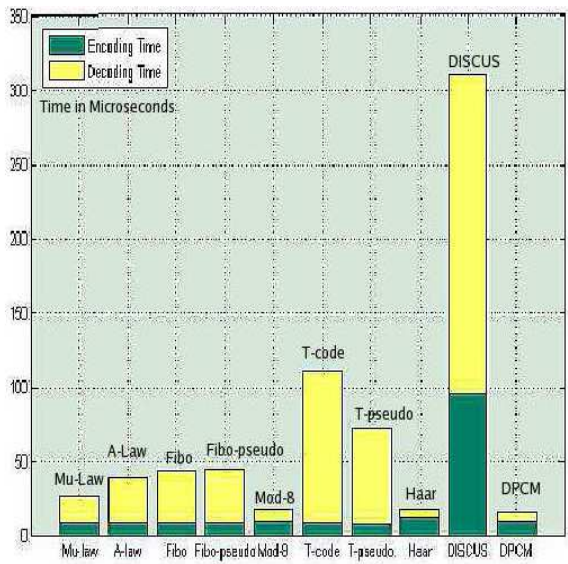}
\caption{Code encoding-decoding times for various codes}
\end{figure}

\begin{figure}[h]
\includegraphics[height=4in,width=3.25in]{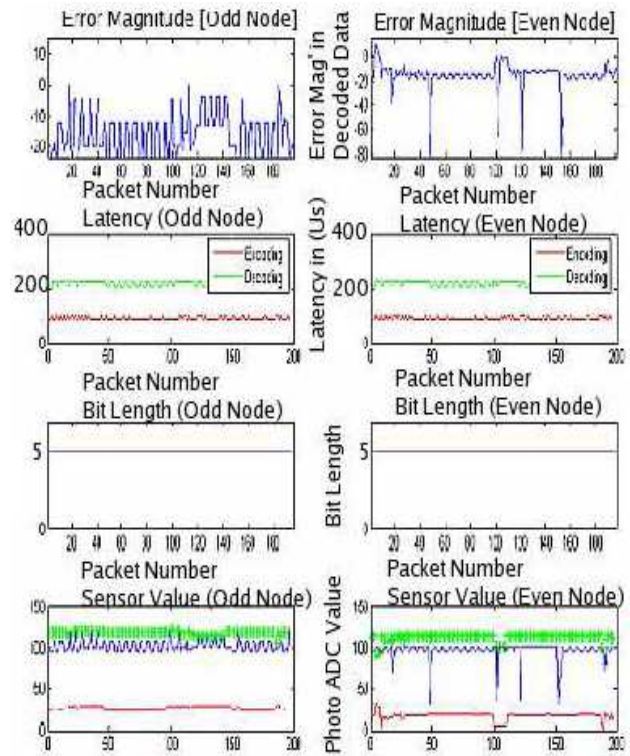}
\caption{DISCUS code performance for odd and even nodes with the real-time,
coded, decoded data, bit rate and error performance with arriving packets.
Graphs on left column correspond to the odd node and right column correspond
to the even node. The Sensor value graph shows coded data in red, raw data in green,
and decoded data in blue.}
\end{figure}

\begin{table}[t]
\caption{Bit rate (Energy Metrics)}
\begin{tabular}{|c|c|c|c|} \hline
Code &Code Avg Bits &
\multicolumn{2}{|c|}{ Total Error} \\ 
&  &  $\mu$ &$\sigma$  \\ \hline
$\mu$-law & 6.5447 & -3.6159 & 2.1767 \\ \hline
A-law & 6.5408 & 8.0306  & 27.7193  \\ \hline
Fibonacci & 10.9765 & 0  & 0  \\ \hline
Fibonacci & 2.2864 & 0 & 0  \\
(pseudo)  &        &   &   \\ \hline
Modulo-8 & 4.9435 & -1.146 & 13.63  \\ \hline
T-code   & 11.5928 & 0  & 0  \\ \hline
T-code & 3.3875 & 0 & 0 \\ 
(pseudo) &  & & \\ \hline
Haar & 6.5242 & -4.0544 & 55.4128 \\ \hline
DISCUS & 5 & -3.7846 & 16.0268 \\ \hline
DPCM & 2.6175 & -21.7274 & 17.3996 \\ \hline
\end{tabular}
\end{table}

Various performance metrics were used in this experiment: \textit{latency}, to show the
encoding and decoding complexity of the code in our implementation,
\textit{energy utilized} in transmission as a function of message length, 
proportional to $430nJ/bit$ (as reported in \cite{energy_calculation}),
\textit{errors} due to decoding, quantization, \textit{size} in bytes of codecs,
and \textit{computational complexity} of the codecs. 

The performance report is presented in [Tables 4,5,6,7].

Graphs for these various parameters are also shown [Fig 5,6,7]. The complete 
analysis for each of the codes is presented like the case of DISCUS code in [Fig 7].

\begin{table}[h]
\caption{Complexity}
\begin{tabular}{|l|l|l|l|l|} \hline
Code & \multicolumn{2}{|c|}{ Complexity } & \multicolumn{2}{|c|}{ Implementation } \\ \hline 
& Enc & Dec & Enc & Dec  \\ \hline
$\mu$-law & $O(1)$ & $O(log(n))$ & LUT & Binary Search  \\ \hline
A-law & $O(1)$ & $O(log(n))$ & LUT & Binary Search  \\ \hline
Fibonacci & $O(1)$ & $O(log(n))$ & LUT & Binary Search  \\ \hline
Modulo-8 & $O(1)$ & $O(log_2n)$ & Compute & Linear Search\\ \hline
T-code& $O(1)$ & $O(n)$ & LUT & Linear Search \\ \hline
T-code& $O(1)$ & $O(n)$  & LUT & Linear Search \\ 
(pseudo) &  &  & & \\ \hline
Haar & $O(1)$ & $O(1)$  & Compute & Compute \\ \hline
DISCUS & $O(n)$ & $O(kn^2)$ &  Compute & Compute \\ \hline
DPCM & O(1) & O(1) & Compute & Compute\\ \hline
\end{tabular}
\end{table}

\begin{table}[h]
\caption{Encoder-Decoder Performance}
\begin{tabular}{|l|l|l|l|l|} \hline
Code & \multicolumn{2}{|c|}{  Encode} & \multicolumn{2}{|c|}{ Decode } \\ \hline
 & $\mu$ & $\sigma$  &  $\mu$ & $\sigma$ \\ \hline
$\mu$-law & 8.6789 & 3.1452 & 17.5325 & 3.6159 \\ \hline
A-law & 8.9789 & 2.7132 & 29.9684 & 2.6435  \\ \hline
Fibonacci & 8.5729 & 2.2267 & 34.4219 & 4.787  \\ \hline
Fibonacci & 8.7805 & 2.4636 & 35.6940 & 2.9272  \\ 
(pseudo) & & & & \\ \hline
Modulo-8 &  9.4545 & 2.5748 & 8.405 & 0.4916 \\ \hline
T-code& 8.7451 & 2.1292  & 101.9477 & 69.6534 \\ \hline
T-code& 7.8000 & 1.7889  & 64.0750 & 25.4980 \\ 
(pseudo) & & & & \\ \hline
Haar & 12.1862 & 3.6064 & 5.8866 & 0.3177 \\ \hline
DISCUS & 95.1515 & 29.3387 & 215.5455 & 4.5695\\ \hline
DPCM & 9.1946 & 2.1936 & 6.1407  & 0.3483 \\ \hline
\end{tabular}
\end{table}

\section{Discussion}

Evidently, the lossless codes have a slightly longer encoding time and 
a longer decoding time compared to the lossy codecs. On the other hand, 
the errors involved in lossless codes are zero while
lossy codecs have significant errors. Similarly the average bit rates for the 
lossless codecs are seen to be higher than that of the lossy coders.
Energy consumption of lossy codecs is much lesser than that of lossless codecs
as it is proportional to the bit rate.

DPCM codes are sensitive to errors due to startup differences.
DPCM errors propagate and affect other decoded samples.
Such codes need to be implemented in a periodic or a per-frame coded manner. 

In the encoding-decoding time graphs the anomalous peaks [Fig 7]  are
the artifacts due to the TinyOS scheduler and the overhead incurred due
to the $\mu$s timer used. This is justified by the constant spike levels seen
in that graph, and the other similar graphs obtained for various codecs which
show a similar anomalies.

In this experiment, the Entropy codes (Huffman) and Universal codes (T-code and Fibonacci codes)
 were mapped on a non-specific source data and associated frequency tables due to which the
 average bit-rates are higher than the optimum (where the probability distribution
of the source model is known). Finding source probability distribution
 is possible in applications like traffic  monitoring and environmental monitoring.
 Design of source codes for those applications are justified by this approach. 
 T-code and  Fibonacci code have a advantage over codes like delta-modulation
 and Haar-wavelet codes which reach the same bit rate or lesser, in
 terms of non-ergodic property and the error-recovery from the unsynchronized
 portions of code. DPCM codes are irreversibly damaged if there is a single error
 in the code for that corresponding frame whereas these entropy-based codes are
 much more reliable for use in sensor networks.

If prior knowledge of the dynamic range of sensor output is known,then
a lossy compression scheme may be implemented by defining a  transfer
function between the sensor values and encoded output at the risk of
introducing errors. 

A LUT can be used to get the encoded value for every sample from the
sensor. Then the problem of source coding and decoding reduces to reading a element 
off the LUT and finding the index corresponding to the element in the LUT.
 As expected, the lossy nature of the encoding schemes attribute to the 
error in decoding values and are only suited for loss tolerant applications
and thresholding/on-off type of applications.

The decoding errors are considerably high for Modulo
codes. The positive aspect of these codes is they require fewer
bits to represent the data samples and hence lesser energy budget. So,
when the perturbation in the input physical variable is small, such
source coding schemes are useful. The encoding latency is not much
of an issue in both these coding schemes. But, unlike T-code, decoding
latency in Modulo code is considerably higher due to complexities
involved in decoding.

In case of Fibonacci codes, for the given link, the reconstruction of
the original values from the sensor is perfect and decoding latency is
well within 40$\mu$s. The complexity involved is that of binary search in
a bisected sorted space. But in this variable bit rate decoding, the
average bit required to represent a sample is pretty high. 
This coding scheme will be valuable when the sensing environment
is highly fluctuating, and near absolute reconstruction of the sensed
values are required. Needless to say, it will take its toll in terms
of transmission energy, as the average bit rate is twice as high as the
other codes.

Since in real world applications we are not able to identify apriori
the probability distribution of the source symbols, adaptive source coding
techniques can be widely used. Schemes like Adaptive-Huffman coding can 
be implemented to achieve the adaptive coding methods, if somehow they
can be implemented within the low-memory constraints and small decoding times,
provided by the sensor platform.

\subsection{Application Specific codes}
A set of applications and specific source-codes
suitable to them, are proposed as a result of the experiments.

\begin{table}[h]
\caption{ Application specific Source Code }
\begin{tabular}{|l|l|}\hline
Source Code &   Application Type \\ \hline
Compander Code &   \\
 A-law, $\mu$-law,DPCM & Fault Tolerant Systems,\\
 &  Event Detection  \\ \hline

Entropy Codes &   \\
T-code, Fibonacci & Mission Critical Systems, \\
 & Data Forwarding \\ \hline

Correlation Codes &   \\
DISCUS, Modulo-N, Haar  & High Sampling Rate systems, \\ 
                        &  Physical Process monitoring \\ \hline
\end{tabular}
\end{table}

\subsection{Limitations of experiment}
The known caveats and limitations of this implementation include:
\begin{enumerate}
\item Timing \& Latency Measurement: These measurements were interrupt driven, and 
have a small but non-negligible overhead in time which was not considered. It is 
expected to have a sub-$\mu$s overhead.
\item Noiseless Channel: No channel errors were considered or reported
from the experiment in lab conditions. This is unlikely in a real world 
scenario.
\item Apriori source codes: Adaptive source codes were neither implemented,
nor tested. Adaptive-Huffman code implementation on a sensor network
 is non-trivial.
\item Application: The sensor network application was prototype measurement
system with no real-application, and only calibration and measurements at
various points of the network.
\item Startup Errors: The first 2 packets in the joint-decoder schemes (Haar,DISCUS) are 
erroneous, due to the implementation.
\end{enumerate}

\subsection{Future work} Dynamic modulation code scaling (DMCS) \cite{code_scaling}
is a viable scheme in a heterogeneous multi-sensing network
where many parameters are measured and processed at various levels of accuracy
and importance. The framework of codes presented here, and their implementations
allow easy implementation of a \textit{dynamically modified} source-coding scheme 
at the network level and thus creating a new kind of optimized network.

Innovations in network coding \cite{nw_code}, can also be explored to provide higher capacity
to sensor networks by cooperative coding.

\section{Conclusion}
This paper presents a study of source codes and their compact
implementation for sensor networks. The results from the analysis of the
data acquired from various source codes used in the network are
mapped to suitable kind of applications in sensor networks according to
the requirements  in error, latency and energy consumption.

%
\bibliographystyle{abbrv}

\end{document}